\documentstyle[amssymb,aps]{revtex}

\begin{document}
\title{{\bf Low Momentum Scattering in the Dirac Equation}}
\author{{\bf Piers Kennedy}$^{\dagger}$, {\bf Norman Dombey}$^{*}$}
\address{Centre for Theoretical Physics, University of Sussex, Brighton BN1 \\
9QJ,~UK\\
email: $^{\dagger }$kapv4@pact.cpes.susx.ac.uk, $^{*}$normand@sussex.ac.uk}
\maketitle

\medskip {\bf Abstract}: It is shown that the amplitude for reflection of a
Dirac particle with arbitrarily low momentum incident on a potential of
finite range is $-1$ and hence the transmission coefficient $T=0$ in
general. If however the potential supports a half-bound state at momentum $k=0$ this
result does not hold. In the case of an asymmetric potential the
transmission coefficient $T$ will be non-zero whilst for a symmetric
potential $T=1.$ Therefore in some circumstances a Dirac particle of
arbitrarily small momentum can tunnel without reflection through a potential
barrier.

\section*{Introduction}

The results for scattering at arbitrary low energy $E$ in one dimension in
the Schr\"{o}dinger equation are well known. If the potential $V(x)$ is
sufficiently well behaved at infinity, then the reflection coefficient at zero
energy is unity and the transmission coefficient is zero \cite{fad} unless
the potential supports a zero energy resonance (a half-bound state). In that
case the transmission coefficient is unity and there is no reflection
provided that the potential is symmetic. Bohm calls this a transmission
resonance \cite{Bohm}. These results have been generalised to asymmetric
potentials \cite{ges}, \cite{SdB}. In this paper we repeat the analysis for
the Dirac equation.\smallskip To some extent this has already been done by
Clemence \cite{clem} in the mathematical literature in his analysis of the
Levinson theorem but we approach the problem as physicists. Our results show
that transmission resonances will occur in the Dirac equation even for the
case where the potential $V(x)$ is everywhere positive and thus represents a
potential barrier.

The potentials $V(x)$ we shall consider are smooth and of finite range$.$ In
non-relativistic systems for such potentials, scattering states with
continuum wave functions have $E\geqslant 0$ whereas bound states with
normalisable wave functions have $E<0$. A half-bound state \cite{Newton} or
zero energy resonance in non-relativistic scattering occurs when the
potential supports a bound state of energy $E=-\kappa ^{2}/2m$ in the limit $%
\kappa \rightarrow 0$: the corresponding wave function thus becomes a
continuum wave function. An example of this is when a square well is
sufficiently deep to just support the first odd bound state: the resulting
wave function describes a non-normalisable half-bound state which
corresponds both to a particle of arbitrarily low energy incident on the
potential from the left and also to a particle of arbitarily low energy
incident from the right.\smallskip

In the relativistic Dirac equation, the notion of half-bound states is more
subtle. For a free Dirac particle, there exists a gap $E\leq |m|$ which
separates the positive and negative energy continuum states: the positive
energy states correspond to particle states and the absence of negative
energy states (hole states) describe anti-particles. On the introduction of
a potential $V(x)$ this gap becomes distorted and bound states now occur
between $E=-m$ and $E=m$. A potential which is attractive to particles and
supports a half-bound state at $E=-m$ or a potential which is attractive to
anti-particles and supports a half-bound state at $E=m$ is called a
supercritical potential: thus the Dirac equation has half-bound states at
both $E=-m$ and $E=m$ in contrast to the Schr\"{o}dinger equation where
these only exist at $E=0$. It follows also that we should talk of zero
momentum resonances in the relativistic case rather than zero energy
resonances.\smallskip

In the following sections we discuss the one-dimensional Dirac equation
using a two-component approach and establish the formalism needed for the
consideration of scattering and bound states. We will then prove that Dirac
particles with energy $E>m$ and arbitrarily small momentum incident on a
potential of finite range will be completely reflected unless the lower
component of a particular wave function vanishes. If the potential supports
a half-bound state at the threshold energy $E=m$ this condition is shown to
be satisfied. In this case there will be a non-zero transmission coefficient
in general whilst for a symmetric potential, there will be a transmission
resonance: the particle will tunnel without reflection. In particular, we
confirm our previous result \cite{DKC} that solutions of the Dirac equation
exist in which a particle of arbitrarily small momentum can tunnel
completely through a potential barrier. In the Appendix we illustrate our
results by considering an asymmetric potential which is soluble analytically.

\section*{The Two-Component Approach}

Following an earlier paper\cite{cdi} we take the gamma matrices $\gamma _{x}$
and $\gamma _{0}$ to be the Pauli matrices $\sigma _{x}$ and $\sigma _{z}$
respectively. Then the Dirac equation for scattering of a particle of energy 
$E$ and momentum $k$ by the potential $V(x)$ is

\begin{equation}
(\sigma _{x}\frac{\partial }{\partial x}-(E-V(x))\sigma _{z}+m)\psi =0
\label{one}
\end{equation}

We write 
\begin{equation}
\psi (x)=\left( 
\begin{array}{c}
f(x) \\ 
g(x)
\end{array}
\right)
\end{equation}
to obtain the coupled differential equations

\begin{mathletters}
\begin{eqnarray}
f^{\,\prime }(x) &=&-\,(E-V(x)+m)\,g(x) \\
g^{\,\prime }(x) &=&\quad \hfill (E-V(x)-m)\,f(x)
\end{eqnarray}
For a free Dirac particle of momentum $k$ the solution is $\psi =\left( 
\begin{array}{c}
A \\ 
B
\end{array}
\right) e^{ikx}$ where $k^{2}=E^{2}-m^{2}$ and 
\end{mathletters}
\begin{equation}
A=\left( \frac{ik}{E-m}\right) B\,=\,i\sqrt{\frac{E+m}{E-m}}\,B\,=\,\left( 
\frac{E+m}{-ik}\right) B
\end{equation}
\noindent Suitable choices for $A$ and $B$ will facilitate future
calculations. For threshold problems where $E\to m$, choosing $B=-ik$ leads
to $A=E+m$ and the free particle wave function $\psi $ can be written apart
from a normalisation factor as 
\begin{equation}
\psi =\left( 
\begin{array}{c}
E+m \\ 
-ik
\end{array}
\right) e^{ikx}  \label{left}
\end{equation}
\noindent It is clear that in this form the top and bottom components do not
simultaneously tend to zero as $E\to m$, $k\to 0$. If on the other hand we
were interested in threshold wave functions where $E\to -m$ then choosing $%
B=E-m$ leads to $A=ik$ and the free wave function can now be written (again
up to normalisation) as 
\begin{equation}
\psi =\left( 
\begin{array}{c}
ik \\ 
E-m
\end{array}
\right) e^{ikx}
\end{equation}
\noindent

\section*{S-Matrix Formalism for the One-Dimensional Dirac Equation}

The S-matrix formalism for scattering in one dimension for the
Schr\"{o}dinger equation is well known and covered in a large number of
texts (e.g \cite{fad}, \cite{CTDL}). The same arguments are applicable for
the Dirac equation in one dimension \cite{bart},\cite{clem} and here we will
summarise a number of the more important results in the context of a
relativistic equation.\smallskip

We adopt the usual formalism for a Dirac particle incident from the left
scattering off the piecewise continuous potential $V(x)$ of finite range
where $V=0$ for $|x|\geq \xi $ where the asymptotic solution $\psi _{l}(x)$
of Eqs. (3) for particles incident from the left with momentum $k$ and
energy $E$ using Eq. (\ref{left}) is

\begin{equation}
\psi _{l}\to \left( 
\begin{array}{c}
E+m \\ 
-ik
\end{array}
\right) e^{ikx}+l(k)\left( 
\begin{array}{c}
E+m \\ 
ik
\end{array}
\right) e^{-ikx},\qquad x\to -\infty  \label{l}
\end{equation}
which defines the (left) reflection amplitude $l(k).$ We can also define the
(left) transmission amplitude $t_{l}(k)$

\begin{equation}
\psi _{l}\to t_{l}(k)\left( 
\begin{array}{c}
E+m \\ 
-ik
\end{array}
\right) e^{ikx},\qquad x\to \infty  \label{t}
\end{equation}
We can similarly define the asymptotic wave function for particles incident
from the right as:

\begin{equation}
\psi _{r}\to t_{r}(k)\left( 
\begin{array}{c}
E+m \\ 
ik
\end{array}
\right) e^{-ikx},\qquad x\to -\infty
\end{equation}
\begin{equation}
\psi _{r}\to \left( 
\begin{array}{c}
E+m \\ 
ik
\end{array}
\right) e^{-ikx}+r(k)\left( 
\begin{array}{c}
E+m \\ 
-ik
\end{array}
\right) e^{ikx},\qquad x\to \infty
\end{equation}
thus defining the right reflection and transmission coefficients $%
r(k),t_{r}(k)$.\smallskip

The left scattering coefficients and the right
coefficients can be simplified further. If we had two independent solutions of
the Dirac equation 
\begin{equation}
\psi _{1}=\left( 
\begin{array}{c}
f_{1}(x) \\ 
g_{1}(x)
\end{array}
\right) \qquad ,\qquad \psi _{2}=\left( 
\begin{array}{c}
f_{2}(x) \\ 
g_{2}(x)
\end{array}
\right)  \label{def}
\end{equation}
then the Wronskian of the solutions $\psi _{1},\psi _{2}$ of the first order
linear differential equations of Eqs. ($3$) defined as \cite{Calogero}:

\begin{equation}
W[\psi _{1},\psi _{2}](x)=f_{1}(x)g_{2}(x)-f_{2}(x)g_{1}(x)
\end{equation}

\noindent would satisfy $W^{\prime }(x)=0$ with the Wronskian $W(x)$ constant
and non-zero. (When $k=0$ it is easy to see that any two solutions are not
independent and $W=0$). We can now evaluate the Wronskian $W(\psi _{l},\psi
_{r})(x)$ as $x\to \pm \infty $ to give 
\begin{equation}
t_{l}(k)=t_{r}(k)=t(k)
\end{equation}
So there is only one transmission coefficient $t(k).\smallskip \smallskip $

The general solution of the Dirac equation $\psi (x)$ can thus be written as
a linear combination of $\psi _{l}$ and $\psi _{r}$: 
\begin{equation}
\psi =A\,\psi _{l}+B\,\psi _{r}
\end{equation}
The asymptotic solutions are now found to be

\begin{equation}
\psi \to A\left( 
\begin{array}{c}
E+m \\ 
-ik
\end{array}
\right) e^{ikx}+\tilde{B}\left( 
\begin{array}{c}
E+m \\ 
ik
\end{array}
\right) e^{-ikx},\qquad x\to -\infty   \label{gen1}
\end{equation}
\begin{equation}
\psi \to \tilde{A}\left( 
\begin{array}{c}
E+m \\ 
-ik
\end{array}
\right) e^{ikx}+B\left( 
\begin{array}{c}
E+m \\ 
ik
\end{array}
\right) e^{-ikx},\qquad x\to \infty   \label{gen2}
\end{equation}
where 
\begin{equation}
\tilde{A}(k)=A\,t(k)+B\,r(k)\quad ,\quad \tilde{B}(k)=A\,l(k)+B\,t(k)
\end{equation}
The coefficients $A$ and $B$ are the amplitudes of the incoming waves for
particles arriving from $x\to -\infty $ and $x\to \infty $ respectively.
Conversely, the coefficients $\tilde{A}$ and $\tilde{B}$ are the
coefficients of the outgoing waves for the transmitted or reflected
particles. We can now introduce the matrix $S(k)$ which allows us to
calculate the outgoing amplitudes in terms of the incoming amplitudes. 
\begin{equation}
\left( 
\begin{array}{c}
\tilde{A} \\ 
\tilde{B}
\end{array}
\right) =S(k)\left( 
\begin{array}{c}
A \\ 
B
\end{array}
\right) \quad \Rightarrow \quad S(k)=\left( 
\begin{array}{cc}
t(k) & r(k) \\ 
l(k) & t(k)
\end{array}
\right)   \label{smat}
\end{equation}
The flux $j$ is given by 
\begin{equation}
j=\overline{\psi }(x)\gamma _{x}\psi (x)=i\,\overline{\psi }(x)\sigma
_{x}\psi =i\,\overline{\psi }(x)\sigma _{z}\sigma _{x}\psi =-\psi ^{\dagger
}(x)\sigma _{y}\psi (x)
\end{equation}
Using equations (\ref{gen1}) and (\ref{gen2}) we consequently find that 
\begin{equation}
\quad 
\begin{array}{cr}
j=2k(E+m)(|A|^{2}-|\tilde{B}|^{2}) & \qquad x\to -\infty  \\[0.3cm]
j=2k(E+m))(|\tilde{A}|^{2}-|B|^{2}) & \qquad x\to \infty 
\end{array}
\end{equation}
The conservation of flux gives us the condition 
\begin{equation}
|A|^{2}+|B|^{2}=|\tilde{A}|^{2}+|\tilde{B}|^{2}
\end{equation}
Also 
\[
|\tilde{A}|^{2}+|\tilde{B}|^{2}=(\tilde{A}^{*}\;\tilde{B}^{*})\left( 
\begin{array}{c}
\tilde{A} \\ 
\tilde{B}
\end{array}
\right) =(A^{*}\;B^{*})S(k)^{\dagger }S(k)\left( 
\begin{array}{c}
A \\ 
B
\end{array}
\right) =|A|^{2}+|B|^{2}
\]
Hence $S(k)$ is a unitary $2\times 2$ matrix. From equation (\ref{smat}),
this imposes the following conditions on the matrix elements of $S(k)$: 
\begin{equation}
T(k)+L(k)=T(k)+R(k)=1  \label{unit}
\end{equation}
\begin{equation}
t(k)r^{*}(k)+t^{*}(k)l(k)=t^{*}(k)r(k)+t(k)l^{*}(k)=0  \label{trl}
\end{equation}
where $T(k)=|t(k)|^{2}$ is the transmission coefficient, $L(k)=|l(k)|^{2}$
is the reflection coefficient for a particle incident from the left and $%
R(k)=|r(k)|^{2}$ is the reflection coefficient for a particle incident from
the right. It also follows that 
\begin{equation}
|l(k)|=|r(k)|
\end{equation}
Additionally, if the potential is symmetric, i.e. $V(x)=V(-x)$ (see ${\bf D}$
below) then $\psi ^{\prime }(x)=\sigma _{z}\psi (-x)$ is also a solution
(this is Eq. (\ref{sym}) below). The asymptotic wave function $\psi ^{\prime
}(x)$ can be found from $\psi (x)$ using equations (\ref{gen1}), (\ref{gen2}%
) by the substitutions $A\leftrightarrow B$ and $\tilde{A}\leftrightarrow 
\tilde{B}$. This implies that $S(k)$ must also be symmetric and consequently 

\begin{equation}
r(k)=l(k).\newline
\label{reql}
\end{equation}
in this case.

The last property we wish to illustrate is the behaviour of the amplitudes $%
t(k)$, $l(k)$ and $r(k)$ at $k=0.$ By taking the complex conjugate of Eqs.
(2-5) with negative momentum $-k$, we see that $\psi _{l,r}^{*}(-k,x)$ has
the same form as $\psi _{l,r}(k,x)$. This in turn implies that 
\begin{equation}
t^{*}(-k)=t(k)\quad ,\quad l^{*}(-k)=l(k)\quad ,\quad r^{*}(-k)=r(k)
\label{star}
\end{equation}

So we see from Eq. (\ref{star}) that all the amplitudes $l(0),r(0),t(0)$ are
real. This will be of importance for the next section. We also have from Eq.
(\ref{trl}) that

\begin{equation}
r(0)=-l(0)\qquad {\bf or}\qquad t(0)=0  \label{spec}
\end{equation}
It follows from Eqs. (\ref{reql}) and (\ref{spec}) that for symmetric
potentials 

\begin{equation}
r(0)=l(0)=0\qquad {\bf or}\qquad t(0)=0  \label{reql0}
\end{equation}
We discuss this further in ${\bf D}$ below.

\section*{Reflection and Transmission Properties at Zero Momentum}

\subsection{The General Case}

Our approach will follow that presented for the Schr\"{o}dinger equation by
Senn \cite{Flugge}. When a Dirac particle is incident from the left
scattering on the potential $V(x)$ of finite range so that $V(x)=0$ for $%
|x|\geq \xi $, the solution $\psi ^{s}$ of Eqs. (3) in Region I $x\leq -\xi $
for particles incident from the left with momentum $k$ and energy $E$ is just

\begin{equation}
\psi ^{s}=\psi _{l}=\left( 
\begin{array}{c}
E+m \\ 
-ik
\end{array}
\right) e^{ikx}+l(k)\left( 
\begin{array}{c}
E+m \\ 
ik
\end{array}
\right) e^{-ikx},\qquad x\leq -\xi
\end{equation}

\smallskip Similarly in Region III $x>\xi $

\begin{equation}
\psi ^{s}=\psi _{l}=t(k)\left( 
\begin{array}{c}
E+m \\ 
-ik
\end{array}
\right) e^{ikx},\qquad x\geq \xi
\end{equation}

For $k\neq 0$ we can define two independent solutions of Eqs. (3) by, for
example,

\begin{equation}
\psi ^{L}=\left( 
\begin{array}{c}
E+m \\ 
-ik
\end{array}
\right) e^{ikx}\qquad x\rightarrow -\infty
\end{equation}

\begin{equation}
\psi ^{R}=\left( 
\begin{array}{c}
E+m \\ 
ik
\end{array}
\right) e^{-ikx}\qquad x\rightarrow \infty
\end{equation}
which represent purely incoming particles from the left and right
respectively. By taking appropriate linear combinations of $\psi ^{L},\psi
^{R}$ and normalising we can choose two new independent solutions of Eqs.
(3) 
\begin{equation}
\psi _{1}=\left( 
\begin{array}{c}
f_{1}(x) \\ 
g_{1}(x)
\end{array}
\right) ,\qquad \psi _{2}=\left( 
\begin{array}{c}
f_{2}(x) \\ 
g_{2}(x)
\end{array}
\right)  \label{ind}
\end{equation}
with the properties

\begin{equation}
g_{1}(-\xi )=0\qquad g_{2}(-\xi )=1\qquad f_{1}(-\xi )=1\qquad f_{2}(-\xi )=0
\label{const}
\end{equation}

Note that the solutions $\psi _{1}$ and $\psi _{2}$ which satisfy Eq. (\ref
{const}) are everywhere real provided that $k$ is real. We can then express
our solution $\psi ^{s}$ in terms of a linear combination of $\psi _{1}$ and 
$\psi _{2}$ for all $x$ and in particular in Region II $|x|\leq \xi $

\begin{equation}
\psi ^{s}=b\left( 
\begin{array}{c}
f_{1}(x) \\ 
g_{1}(x)
\end{array}
\right) +c\left( 
\begin{array}{c}
f_{2}(x) \\ 
g_{2}(x)
\end{array}
\right) \qquad -\xi \leq x\leq \xi .  \label{scatt}
\end{equation}

\noindent

We can evaluate the Wronskian of the solutions $\psi _{1},\psi _{2}$ is
constant at the point $x=-\xi $ to give

\[
W[\psi _{1},\psi _{2}]=W(-\xi )=f_{1}(-\xi )g_{2}(-\xi )-f_{2}(-\xi )g(-\xi
)=1 
\]
thus confirming that the solutions $\psi _{1},\psi _{2}$ are independent for 
$k\neq 0.\smallskip $

\smallskip The wave function $\psi ^{s}(x)$ must be continuous at $x=-\xi $
and $x=\xi $. The overlap between Regions I and II and between II and III
then give the following boundary conditions:

\begin{mathletters}
\begin{eqnarray}
(E+m)(e^{-ik\xi }+l(k)e^{ik\xi })\; &=&\;b \\
-ik(e^{-ik\xi }-l(k)e^{ik\xi })\; &=&\;c \\
(E+m)t(k)e^{ik\xi }\; &=&\;bf_{1}(\xi )+cf_{2}(\xi ) \\
-ikt(k)e^{ik\xi }\; &=&\;bg_{1}(\xi )+cg_{2}(\xi )
\end{eqnarray}

For simplicity, write $\alpha _{i}=f_{i}(\xi )$ and $\beta _{i}=g_{i}(\xi )$%
, so that the last two equations become

\end{mathletters}
\begin{mathletters}
\begin{eqnarray}
(E+m)t(k)e^{ik\xi }\; &=&\;b\alpha _{1}+c\alpha _{2} \\
-ikt(k)e^{ik\xi }\; &=&\;b\beta _{1}+c\beta _{2}
\end{eqnarray}

Note that $b$ and $c$ are dependent on $k$ as are $\alpha _{i}$ and $\beta
_{i}.$ Eliminating $t$, $b$ and $c$ then re-arranging to solve for $l$ gives

\end{mathletters}
\begin{equation}
l(k)=\left( \frac{k^{2}\alpha _{2}+(E+m)^{2}\beta _{1}+ik(E+m)(\alpha
_{1}-\beta _{2})}{k^{2}\alpha _{2}-(E+m)^{2}\beta _{1}-ik(E+m)(\alpha
_{1}+\beta _{2})}\right) e^{-2ik\xi }  \label{refl}
\end{equation}

Similarly $t$ can be found to be

\begin{equation}
t(k)=\frac{-2ik(E+m)(\alpha _{1}\beta _{2}-\alpha _{2}\beta _{1})}{%
k^{2}\alpha _{2}-(E+m)^{2}\beta _{1}-ik(E+m)(\alpha _{1}+\beta _{2})}%
e^{-2ik\xi }  \label{trans}
\end{equation}
We can then use the relation

\begin{equation}
W(\xi )=f_{1}(\xi )g_{2}(\xi )-f_{2}(\xi )g_{1}(\xi )=\alpha _{1}\beta
_{2}-\alpha _{2}\beta _{1}=1
\end{equation}
to simplify Eq (\ref{trans}). It is a straightforward exercise to verify
that Eqs. (\ref{refl}, \ref{trans}) satisfy the unitarity condition Eq. (\ref
{unit}).\smallskip

We can now discuss the limit as $k\rightarrow 0.$ It is apparent from Eqs. (%
\ref{refl}, \ref{trans}) that provided $\beta _{1}(0)\neq 0$ the limit $E\to
m$, $k\to 0$ gives

\begin{equation}
l(0)=r(0)=-1\qquad t(0)=0  \label{gener}
\end{equation}
so that the reflection coefficients $L(0)=R(0)=1$ and the transmission
coefficient $T(0)=0$ . These results in the general case agree with those
for the Schr\"{o}dinger equation \cite{ges}, \cite{SdB}.\smallskip

Using Eqs. (33a,33b) it can be seen that $b(0)=c(0)=0$ in the $k=0$ limit
and therefore from Eq.(\ref{scatt}) the wave function vanishes identically
for all $x$. Thus the only physical solution of the Dirac equation $(1)$ for 
$k=0$ is the solution

\begin{equation}
\psi (x,k=0)=0  \label{null}
\end{equation}
unless the potential has special properties which we investigate in the next
section.\smallskip

It should also be noted that as $f_{i}(x)$ and $g_{i}(x)$ are real at $x=\xi 
$, the quantities $\alpha _{i}(k)$ and $\beta _{i}(k)$ are also real. Hence
from Eq. (\ref{trans}) as $k\to 0$, $t(k)$ is pure imaginary as it
approaches zero in agreement with the Levinson theorem \cite{clem} for the
Dirac equation provided $\beta _{1}(0)\neq 0.$

\subsection{The special case $\beta _{1}(0)=0$}

If we return to Equations (33b,34b) we see that as $k\rightarrow 0$ we must
have

\[
c(0)=0\qquad b(0)\beta _{1}(0)+c(0)\beta _{2}(0)=0 
\]

so

\begin{equation}
b(0)\beta _{1}(0)=0  \label{cond}
\end{equation}
Furthermore since $c(0)=0,$ we must have from Eq. (34a)

\begin{equation}
2mt(0)=b(0)\alpha _{1}(0)  \label{ggg}
\end{equation}

When $b(0)=0$ as well as $c(0)=0$ we obtain the general case already
discussed. If $\beta _{1}(0)=0$, however, then as we approach the limit $%
E\to m$, $k\to 0$ the reflection amplitude $l(k)$ does not satisfy $l(0)=-1$
and so the wave function $\psi (x,k=0)\neq 0$. In this case we therefore
have non-trivial solutions of the Dirac equation at $k=0.$ This implies that
transmission coefficient $t(0)$ will be non-zero in this limit as will $%
\alpha _{1}(0).\smallskip $

For $k\rightarrow 0$ with $\beta _{1}(0)=0$ we can write $\beta
_{1}(k)=k\beta _{1}^{^{\prime }}(0)$ $.$ So from Eq. (\ref{refl}) 
\begin{equation}
\;l(0)=\lim_{k\to 0}\frac{\beta _{2}-\alpha _{1}+2m\beta _{1}^{^{\prime
}}(0)i}{\beta _{2}+\alpha _{1}+2m\beta _{1}^{^{\prime }}(0)i}
\end{equation}
As $k$ is arbitrarily small (and not actually equal to zero), the Wronskian, 
$W=\alpha _{1}\beta _{2}=1+O(k)$ so $\beta _{2}=1/\alpha _{1}+O(k)$ and in
the limit $k\rightarrow 0$ we have 
\begin{equation}
l(0)=\frac{1-\alpha _{1}^{2}(0)+2mi\alpha _{1}(0)\beta _{1}^{^{\prime }}(0)}{%
1+\alpha _{1}^{2}(0)-2mi\alpha _{1}(0)\beta _{1}^{^{\prime }}(0)}
\label{fff}
\end{equation}
We know however from Eq. (\ref{star}) that $l(0)$ must be real. From Eq.(\ref
{fff}) this means that either $\beta _{1}^{^{\prime }}(0)=0$ or $\alpha
_{1}(0)=0.$ But since we are considering the non-trivial case where $\psi
(x,k=0)\neq 0$ (and hence we expect that $t(0)\neq 0)$ we do not want $%
\alpha _{1}(0)=0$ since from Eq. (\ref{ggg}) that would imply that $t(0)=0.$
Thus we would like to be able to show that

\begin{equation}
\beta _{1}^{^{\prime }}(0)=0  \label{lam}
\end{equation}
and $\beta _{1}(k)=O(k^{2}).$ This is not difficult to demonstrate using an
argument of Lin's \cite{lin}: the wave function $\psi _{1}$of Eq. (\ref{def}%
) is a solution of Eqs. $(3)$ subject to the $k$-independent boundary
conditions given by Eq. \ref{const}). So its lower component

\begin{equation}
g_{1}(x,k)=g_{1}(x,E)
\end{equation}
since the Dirac equation $(3)$ involves $E$ explicitly not $k.$ It follows
that

\begin{equation}
\beta _{1}=\beta _{1}(E)
\end{equation}
which requires $\beta _{1}$ to be an even function of $k$ and in particular
that as $E=\sqrt{m^{2}+k^{2}}$

\begin{equation}
\frac{d\beta _{1}(k)}{dk}=\frac{d\beta _{1}(E)}{dE}\frac{dE}{dk}=\frac{k}{E}%
\frac{d\beta _{1}(E)}{dE}=0
\end{equation}
at $k=0$ in agreement with Eq. (\ref{lam}).\smallskip \smallskip

This gives the final result for the reflection amplitudes in the special
case when $\beta _{1}(0)=0$:

\begin{equation}
l(0)=-r(0)=\frac{1-\alpha _{1}^{2}(0)}{1+\alpha _{1}^{2}(0)}  \label{refl2}
\end{equation}
and for the corresponding transmission amplitude from Eq. (\ref{trans}):

\begin{equation}
t(0)=\frac{2\alpha _{1}(0)}{1+\alpha _{1}^{2}(0)}  \label{tr2}
\end{equation}
These results agree with those obtained by Clemence \cite{clem}.

\subsection{Half-Bound State}

We will now show that if the potential were to support a bound state in the
limit $E=m$ then $\beta _{1}(0)=0$ so the scattering wave function will not
vanish in the limit $k\rightarrow 0.$ For an asymmetric potential the
following bound state wave function is appropriate for $|x|\geq \xi $: 
\begin{equation}
\begin{array}{lcr}
Region\,I\qquad & \psi ^{b}=s\left( 
\begin{array}{c}
E+m \\ 
-\kappa
\end{array}
\right) e^{\kappa x} & \qquad x\leq -\xi \\[0.5cm] 
Region\,III\qquad & \psi ^{b}=s^{\prime }\left( 
\begin{array}{c}
E+m \\ 
\kappa
\end{array}
\right) e^{-\kappa x} & \qquad x\geq \xi
\end{array}
\label{bound}
\end{equation}
If the potential is such that the wave function $\psi ^{b}$ possesses a
well-defined non-zero limit as $E\to m$, $\kappa \to 0$, then the wave
function for $|x|\geq \xi $ in this limit is just proportional to 
\begin{equation}
\left( 
\begin{array}{c}
2m \\ 
0
\end{array}
\right)  \label{half}
\end{equation}
albeit with different constants of proportionality $s,s^{\prime }$ on the
left and right$.$ It is clear that a wave function of this form is
non-normalisable and forms part of the continuum.\smallskip

The scattering solutions $\psi ^{s}$ which tend to the solutions Eq. (\ref
{half}) in the zero-momentum limit will therefore have a lower component
which vanishes for sufficiently large $\left| x\right| $. From Eq. (\ref
{scatt}) this implies that at $x=\xi $

\begin{equation}
b(0)\beta _{1}(0)+c(0)\beta _{2}(0)=0
\end{equation}
while at $x=-\xi $ using Eq. (\ref{const}) we have $c(0)=0.$ Since $\psi
^{s}(k=0)$ is not zero for a half-bound state, $b(0)\neq 0$ and hence

\begin{equation}
\beta _{1}(0)=0
\end{equation}

An example of a half bound state in an asymmetric potential is given in the
Appendix together with an explicit demonstration that $\beta _{1}(k)$ is of
order $k^{2}$ for small $k$ when the condition $\beta _{1}(0)=0$ holds .

\subsection{Symmetric Potentials}

When the potential is symmetric so that $V(x)=V(-x)$ we can find more
stringent conditions on $l(0),$ $r(0)$ and $t(0)$. In the two-component
approach, the behaviour of the wave function under the parity transformation 
$x\to -x$ is given by: 
\begin{equation}
\psi ^{\,\prime }(-x,t)=\sigma _{z}\psi (x,t)  \label{sym}
\end{equation}
It follows that we can define an even wave function $\psi _{+}(x)$ under
parity as one with an even top component and an odd bottom component whereas
an odd wave function $\psi _{-}(x)$ has an odd top component and an even
bottom component. The wave function $\psi ^{b}$ for the bound state given in
Eq. (\ref{bound}) must now be either an even solution $\psi _{+}$ or an odd
solution $\psi _{-}.$ First let us assume that it is even.\smallskip 

Then in the limit of a half-bound state at $E=m$, ($\kappa \to 0$) the
solution remains even. As $k\to 0$ the scattering solution $\psi ^{s}$ will
also be even. Thus from Eqs. (\ref{l}, \ref{t}) we have

\begin{equation}
1+l(0)=t(0)  \label{e}
\end{equation}
From the unitarity relation we also know that

\[
l(0)^{2}+t(0)^{2}=1=l(0)^{2}+(1+l(0))^{2} 
\]
therefore

\begin{equation}
l(0)^{2}+l(0)=0
\end{equation}
So either $l(0)=0$ or $l(0)=-1$ in agreement with Eq.(\ref{reql0}). We know
that $l(0)\neq -1$ as $\psi ^{s}(k=0)\neq 0.$ Hence

\begin{equation}
l(0)=0  \label{a}
\end{equation}
and the transmission coefficient

\begin{equation}
T(0)=1  \label{b}
\end{equation}
Using Eq (\ref{tr2}) we see that for an even half-bound state we must have $%
\alpha _{1}(0)=1$ while for an odd half bound state we have $\alpha
_{1}(0)=-1.$

So we obtain the result that when a symmetric potential supports a half
bound state, a transmission resonance $T=1$ occurs for an incident particle
with arbitrarily small momentum.This agrees with our previous result for
reflectionless scattering by a repulsive potential $V(x)$ where its
attractive counterpart $U(x)=-V(x)$ is supercritical \cite{DKC}, that is to
say $U(x)$ has a half-bound state at $E=-m.$ To see this note that  Eqs. (3)
are invariant under the (charge conjugation) transformation

\begin{equation}
E\rightarrow -E\quad V\rightarrow -V\quad f\rightarrow g\quad g\rightarrow f
\end{equation}
so it follows that $V(x)$ has a half-bound state at $E=m$ when $U(x)$ has a
half-bound state at $E=-m.$

\section*{Discussion}

We have now generalised the results for scattering in one dimension in the
Schr\"{o}dinger equation to the Dirac equation as we intended. But we are
physicists not mathematicians: consequently our results are not yet as
complete as those proven for the Schr\"{o}dinger equation. Clemence \cite
{clem}, however, has shown that the class of potentials for which our
results are true in the Dirac equation can be extended to include potentials
which do not vanish for $|x|\geq \xi .$ His results require the potentials $%
V(x)$ to satisfy 

\begin{equation}
\int_{-\infty }^{\infty }(1+|x|)|V(x)|dx<\infty
\end{equation}

As stated in the Introduction a half-bound state at $E=m$ can arise in two
ways in the Dirac equation. These can most easily be distinguished by the
examples of an attractive well for which $V(x)\leq 0$ and a repulsive
barrier for which $V(x)\geq 0$, although it may be more difficult to
characterise which is which for a complicated potential. In the case of an
attractive potential a half-bound state with $E=m$ corresponds to a
non-relativistic zero energy resonance. For example in the case of a square
well $V(x)=-V_{0},|x|\leq a,\,V(x)=0$ elsewhere one occurs at the threshold
for the first odd state $V_{0}=\pi ^{2}/2ma^{2}.$ In the case of a repulsive
potential a half-bound state occurs as we have just seen when the
corresponding attractive potential $U(x)=-V(x)$ is supercritical. For the
square barrier $V(x)=V_{0},|x|\leq a,V(x)=0$ elsewhere supercriticality
first occurs when $V_{0}=m+\sqrt{m^{2}+\pi ^{2}/4a^{2}}$ \cite{cdi}. Note
that $V_{0}>2m$ before supercriticality can occur.\smallskip

Over 70 years ago Klein \cite{klein} discovered that a Dirac particle could
tunnel through a potential barrier $V$ with $V>2m$. In this paper we have
confirmed that tunnelling will always occur in the Dirac equation if a
potential barrier $V(x)$ of short range is strong enough so that $U(x)=-V(x)$
is supercritical. The generic phenomenon whereby fermions can tunnel through
barriers without exponential suppression we have called ``Klein Tunnelling'' 
\cite{cd}. Even strong long range repulsive potentials in the Dirac equation
seem to have this property: in three dimensions Hall and one of us (ND) \cite
{hall} have shown that Klein tunnelling is also associated with
supercriticality for Coulomb potentials.

\medskip {\it Acknowledgments}~We would like to thank Peter Bushell and Alex
Sobolev for their help.

\section{Appendix}

In order to illustrate scattering off an asymmetric potential we shall
consider one of the few examples which can be solved analytically. We shall
use a double delta potential barrier which comprises two unequal Dirac delta
functions: 
\begin{equation}
V(x)=\lambda \,\delta (x)+\mu \,\delta (x-a)  \label{doub}
\end{equation}
where $\lambda \neq \mu $ and $\lambda $, $\mu >0$.

\subsection{Scattering Coefficients}

The wave function for $x<0$ is 
\begin{equation}
\psi (x)=\left( 
\begin{array}{c}
E+m \\ 
-ik
\end{array}
\right) e^{ikx}+l\left( 
\begin{array}{c}
E+m \\ 
ik
\end{array}
\right) e^{-ikx}
\end{equation}
while for $0<x<a$ it is 
\begin{equation}
\psi (x)=\alpha \left( 
\begin{array}{c}
E+m \\ 
-ik
\end{array}
\right) e^{ikx}+\beta \left( 
\begin{array}{c}
E+m \\ 
ik
\end{array}
\right) e^{-ikx}
\end{equation}
and for $x>a$ 
\begin{equation}
\psi (x)=t\left( 
\begin{array}{c}
E+m \\ 
-ik
\end{array}
\right) e^{ikx}
\end{equation}
The discontinuity condition on $\psi (x)$ at the first barrier at $x=0_{\pm
} $ is \cite{cdi} 
\begin{equation}
\psi (0_{+})=e^{i\lambda \sigma _{2}}\psi (0_{-})=\left( 
\begin{array}{rc}
cos\,\lambda & sin\,\lambda \\ 
-sin\,\lambda & cos\,\lambda
\end{array}
\right) \psi (0_{-})
\end{equation}
The second discontinuity condition at $x=a_{\pm }$ is derived by replacing $%
0_{\pm }$ with $a_{\pm }$ and $\lambda $ with $\mu $.\newline
\smallskip

The reflection and transmission amplitudes $l$ and $t$ can then be
calculated to give: 
\begin{equation}
l=-\frac{imk(cos\,\mu \,sin\,\lambda +e^{2iak}cos\,\lambda \,sin\,\mu
)+mE(e^{2iak}-1)sin\,\lambda \,sin\,\mu }{m^{2}(e^{2iak}-1)sin\,\lambda
\,sin\,\mu +k^{2}cos(\lambda +\mu )+iEk\,sin(\lambda +\mu )}
\end{equation}
and 
\begin{equation}
t=\frac{k^{2}}{m^{2}(e^{2iak}-1)sin\,\lambda \,sin\,\mu +k^{2}cos(\lambda
+\mu )+iEk\,sin(\lambda +\mu )}
\end{equation}
Using $E=\sqrt{k^{2}+m^{2}}$ we can write for small $k$%
\begin{equation}
l=\frac{-im(sin(\lambda +\mu )+2ma\,sin\,\lambda \,sin\,\mu
)+2amk(am\,sin\,\lambda \,sin\,\mu +cos\,\lambda \,sin\,\mu )+O(k^{2})}{%
im(sin(\lambda +\mu )+2ma\,sin\,\lambda \,sin\,\mu )+k(cos(\lambda +\mu
)-2a^{2}m^{2}sin\,\lambda \,sin\,\mu )+O(k^{2})}  \label{newl}
\end{equation}
\begin{equation}
t=\frac{k}{im(sin(\lambda +\mu )+2ma\,sin\,\lambda \,sin\,\mu
)+k(cos(\lambda +\mu )-2a^{2}m^{2}sin\,\lambda \,sin\,\mu )+O(k^{2})}
\label{newt}
\end{equation}

From Eqs. (\ref{newl}, \ref{newt}) it is easy to see that in general as $%
k\to 0$

\begin{equation}
l\to -1\quad t\to 0
\end{equation}
in agreement with Eq. (\ref{gener}). If however 
\begin{equation}
sin(\lambda +\mu )+2ma\,sin\,\lambda \,sin\,\mu =0  \label{exc}
\end{equation}
then 
\begin{equation}
l\to -\frac{am\,sin(\lambda -\mu )}{cos(\lambda +\mu )+am\,sin(\lambda +\mu )%
}  \label{reflex}
\end{equation}
and 
\begin{equation}
t\to \frac{1}{cos(\lambda +\mu )-2a^{2}m^{2}\,sin\,\lambda \,sin\,\mu }
\label{transex}
\end{equation}
It is easy to show that $l$ and $t$ given above do indeed satisfy

\[
\left| l\right| ^{2}+\left| t\right| ^{2}=1 
\]
provided that $sin(\lambda +\mu )+2ma\,sin\,\lambda \,sin\,\mu =0.$

\subsection{Exceptional Case}

The exceptional case in the proof above occurs when $\beta _{1}(0)=0.$ We
shall therefore calculate $\alpha _{1}(0)$ and $\beta _{1}(0)$ for the
double delta potential. From Eq. (\ref{const}) we consider the solution of
the Dirac equation which takes the values $\left( 
\begin{array}{l}
1 \\ 
0
\end{array}
\right) $ at $x=-\xi $. The wave function $\psi (x)$ for $x<0$ thus has the
form 
\begin{equation}
(E+m)\psi (x)=\left( 
\begin{array}{c}
(E+m)\,cos\,k(x+\xi ) \\ 
k\,sin\,k(x+\xi )
\end{array}
\right)
\end{equation}
while for $0<x<a$ it is 
\begin{equation}
(E+m)\psi (x)=\gamma \left( 
\begin{array}{c}
(E+m)\,cos\,kx \\ 
k\,sin\,kx
\end{array}
\right) +\delta \left( 
\begin{array}{c}
(E+m)\,sin\,kx \\ 
-k\,cos\,kx
\end{array}
\right)
\end{equation}
and for $x>a$ we can write 
\begin{equation}
(E+m)\psi (x)=\sigma \left( 
\begin{array}{c}
(E+m)\,cos\,k(x-\xi ) \\ 
k\,sin\,k(x-\xi )
\end{array}
\right) +\tau \left( 
\begin{array}{c}
(E+m)\,sin\,k(x-\xi ) \\ 
-k\,cos\,k(x-\xi )
\end{array}
\right)
\end{equation}

So at $x=\xi $ we see that

\begin{equation}
\alpha _{1}(k)=\sigma \quad \beta _{1}(k)=-k\tau /(E+m)
\end{equation}
For small $k$ we calculate from the discontinuity conditions that 
\begin{equation}
\sigma =\left[ cos(\lambda +\mu )+2m\xi \,sin(\lambda +\mu )-2ma\,sin\,\mu
\,cos\,\lambda +4am^{2}(\xi -a)sin\,\mu \,sin\,\lambda \right] +O(k^{2})
\end{equation}
and 
\begin{equation}
k\tau =2m\left[ (sin(\lambda +\mu )+2am\,sin\,\lambda \,sin\,\mu )\right]
+O(k^{2})
\end{equation}
Note that neither $\sigma $ nor $k\tau $ has any term of order $k.$ As $k\to
0$ we obtain

\[
\beta _{1}(0)=-\left[ (sin(\lambda +\mu )+2am\,sin\,\lambda \,sin\,\mu
)\right] \quad \beta _{1}^{^{\prime }}(0)=0 
\]
so the exceptional case given by Eq. (\ref{exc}) above indeed satisfies $%
\beta _{1}(0)=0.$ Furthermore when $\beta _{1}(0)=0$ it is easy to see that

\begin{equation}
\alpha _{1}(0)=cos(\lambda +\mu )-2am\,sin\,\mu \,(cos\,\lambda
+2am\,sin\,\lambda )
\end{equation}

From Eq. (\ref{tr2}) above the transmission coefficient in the exceptional
case when

\[
\beta _{1}(0)=-\left[ (sin(\lambda +\mu )+2am\,sin\,\lambda \,sin\,\mu
)\right] =0 
\]
can be expressed in terms of $\alpha _{1}(0)$%
\begin{equation}
t=\frac{2\alpha _{1}(0)}{1+\alpha _{1}^{2}(0)}
\end{equation}
After some tedious manipulation we find that 
\begin{eqnarray}
1+[\alpha _{1}(0)]^{2} &=&2(1+2am\,sin\,\lambda \,cos\,\lambda
+2a^{2}m^{2}\,sin^{2}\lambda )  \nonumber \\
&=&2(cos(\lambda +\mu )+2am\,sin\,\lambda \,cos\,\mu )(cos(\lambda +\mu
)-2a^{2}m^{2}\,sin\,\lambda \,sin\,\mu )  \nonumber \\
&=&2\alpha _{1}(0)(cos(\lambda +\mu )-2a^{2}m^{2}\,sin\,\lambda \,sin\,\mu )
\end{eqnarray}
So 
\begin{equation}
t=\frac{2\alpha _{1}(0)}{2\alpha _{1}(0)[cos(\lambda +\mu
)-2a^{2}m^{2}\,sin\,\lambda \,sin\,\mu ]}
\end{equation}
in agreement with Eq. \ref{transex}). Similarly it can be shown that Eq. (%
\ref{reflex}) for the reflection coefficient agrees with Eq. (\ref{refl2}).

\subsection{Bound States}

Let us now consider the asymmetric potential well 
\begin{equation}
U(x)=-V(x)=-\lambda \,\delta (x)-\mu \,\delta (x-a)
\end{equation}
This will have bound states with a wave function for $x<0$ of the form 
\begin{equation}
\psi (x)=\left( 
\begin{array}{c}
-\kappa \\ 
m-E
\end{array}
\right) e^{\kappa x}
\end{equation}
while for $0<x<a$ 
\begin{equation}
\psi (x)=\gamma \left( 
\begin{array}{c}
-\kappa \\ 
m-E
\end{array}
\right) e^{\kappa x}+\delta \left( 
\begin{array}{c}
\kappa \\ 
m-E
\end{array}
\right) e^{-\kappa x}
\end{equation}
and for $x>a$ 
\begin{equation}
\psi (x)=s\left( 
\begin{array}{c}
\kappa \\ 
m-E
\end{array}
\right) e^{-\kappa x}
\end{equation}
The discontinuity condition for the first delta well is 
\begin{equation}
\psi (0_{+})=e^{-i\sigma _{2}\lambda }\psi (0_{-})=\left( 
\begin{array}{rc}
cos\,\lambda & -sin\,\lambda \\ 
sin\,\lambda & cos\,\lambda
\end{array}
\right) \psi (0_{-})
\end{equation}
Note that this differs from the condition for barriers only in that $\lambda
\to -\lambda $. The second discontinuity condition follows with $0_{\pm
}\rightarrow a_{\pm }$ and $\lambda \rightarrow \mu $. This leads to the
following four equations:

\begin{mathletters}
\begin{eqnarray}
\kappa (-\gamma +\delta )cos\,\lambda +(m-E)(\gamma +\delta )sin\,\lambda
&=&-\kappa \\
-(m-E)(\gamma +\delta )cos\,\lambda -\kappa (-\gamma +\delta )sin\,\lambda
&=&m-E \\
\kappa (-\gamma e^{a\kappa }+\delta e^{-a\kappa })cos\,\mu +(m-E)(\gamma
e^{a\kappa }+\delta e^{-a\kappa })sin\,\mu &=&s\kappa e^{-a\kappa } \\
(m-E)(\gamma e^{a\kappa }+\delta e^{-a\kappa })cos\,\mu +\kappa (-\gamma
e^{a\kappa }+\delta e^{-a\kappa })sin\,\mu &=&s(m-E)e^{-a\kappa }
\end{eqnarray}
$\gamma $ and $\delta $ can be found from the first two equations ($88a,88b)$
to be: 
\end{mathletters}
\begin{eqnarray}
\gamma &=&-\frac{Esin\,\lambda -\kappa cos\,\lambda }{\kappa }  \nonumber \\
\delta &=&-\frac{msin\,\lambda }{\kappa }
\end{eqnarray}
Eliminating $s$ from $(88c,88d)$ leads to: 
\begin{equation}
\gamma e^{2a\kappa }(\kappa cos\,\mu -Esin\,\mu )+\delta m\sin \,\mu =0
\end{equation}
We thus obtain: 
\begin{equation}
e^{2a\kappa }(\kappa cos\,\lambda -Esin\,\lambda )(\kappa cos\,\mu
-Esin\,\mu )-m^{2}sin\,\lambda \,sin\,\mu =0
\end{equation}
Re-arranging gives 
\begin{equation}
sin(\lambda +\mu )=\frac{\kappa ^{2}e^{2a\kappa }cos\,\lambda \,cos\,\mu
+(e^{2a\kappa }E^{2}-m^{2})sin\,\lambda \,sin\,\mu }{E\kappa e^{2a\kappa }}
\end{equation}
At supercriticality $E=-m,\kappa =0$ giving 
\begin{equation}
sin(\lambda +\mu )+2ma\,sin\,\lambda \,sin\,\mu =0
\end{equation}
in agreement with the exceptional condition $\beta _{1}(0)=0.$ \smallskip

When $\lambda =\mu $ we obtain a symmetric potential. If $sin(\lambda +\mu
)+2ma\,sin\,\lambda \,sin\,\mu =0$ then either $\sin \lambda =0$ and $\alpha
_{1}(0)=1$ or $\tan \lambda =-1/ma$ and $\alpha _{1}(0)=-1.$ In both cases
the transmission coefficient $T=1$ in agreement with our previous result 
\cite{DKC} for supercritical symmetric potentials.

\end{document}